\documentclass[12pt]{article}
\usepackage{amsmath}
\def\a{\alpha}
\def\b{\beta}

\def\d{\delta}

\def\be{\begin{equation}}
\def\ee{\end{equation}}
\def\arr{\begin{array}{rll}}
\def\ea{\end{array}}
\def\bea{\begin{eqnarray}}
\def\eea{\end{eqnarray}}

\def\N2{$N{=}2$}

\def\>{\rangle}
\def\<{\langle}
\def\+{\dagger}
\def\={\ =\ }

\begin{document}
\renewcommand{\thefootnote}{\fnsymbol{footnote}}
\begin{titlepage}
\setcounter{page}{0}
\begin{flushright}
LMP-TPU--9/08  \\
\end{flushright}
\vskip 1cm
\begin{center}
{\LARGE\bf Remark on quantum mechanics with }\\
\vskip 0.5cm
{\LARGE\bf conformal Galilean symmetry }\\
\vskip 2cm
$
\textrm{\Large A.V. Galajinsky\ }
$
\vskip 0.7cm
{\it
Laboratory of Mathematical Physics, Tomsk Polytechnic University, \\
634050 Tomsk, Lenin Ave. 30, Russian Federation} \\
{Email: galajin@mph.phtd.tpu.edu.ru}

\end{center}
\vskip 1cm
\begin{abstract} \noindent

\end{abstract}
Conformal Galilei algebra contains $so(1,2)$ subalgebra which is the conformal algebra in
one dimension. In this note we generalize methods previously
developed for one--dimensional many--body
systems and construct a unitary map relating a quantum mechanics invariant under
the conformal Galilean transformations to a set of decoupled particles for which the
same symmetry is realized in a nonlocal way. Possible applications of the map are
discussed.

\end{titlepage}

\renewcommand{\thefootnote}{\arabic{footnote}}
\setcounter{footnote}0

Conformal Galilean symmetry has been extensively studied
in the past in the context of quantum mechanics, statistical physics and  nonrelativistic
field theory (see e.g. \cite{nied}--\cite{son} and references therein).
More recently, various proposals for a nonrelativistic version of the AdS/CFT
correspondence \cite{son1}--\cite{hor} stimulated a
renewed interest in the conformal Galilean invariance.

The generators of time translation, dilatation and special conformal
transformation entering the Galilei algebra
obey the commutation relations of $so(1,2)$ which is
the conformal algebra in one dimension. By this reason it is natural to
expect that quantum mechanics invariant under the conformal Galilean
transformations may share some interesting features with one--dimensional
conformal
many--body systems. The most prominent example of such a one--dimensional
system seems to be the Calogero model \cite{calo1}.

An interesting peculiarity of the Calogero model with the oscillator potential
is that it can be transformed into a set of decoupled oscillators by applying an
appropriate similarity transformation \cite{gur,glp},
the fact anticipated by Calogero in \cite{calo}.
In particular, this explains
why the spectra of the Calogero model and the decoupled oscillators are so
alike.

When the oscillator potential is absent the Calogero model exhibits conformal
invariance. In \cite{glp,pol} a unitary transformation to a set of decoupled
particles was constructed. The simplification
in the dynamics was achieved at the price of a nonlocal realization of the
full conformal algebra in the Hilbert space \cite{glp}.

The similarity transformation to decoupled particles
suggested a simple and straightforward method of building the complete set of
eigenstates for the Calogero model \cite{gur}. It also provided
an efficient means for constructing various $N=2$ and $N=4$ superconformal
many--body models in one dimension \cite{glp,glp1,glp2,ol}. An
elegant geometric
interpretation
of the similarity transformation as the inversion of the Klein model of the Lobachevsky
plane was proposed in \cite{arm}.

The purpose of this note is to generalize the results previously obtained for
one--dimensio\-nal systems \cite{glp} to higher dimensions, i.e. to the
case of the conformal Galilei algebra. Our analysis also
covers the time--dependent parts of the conformal generators which
were disregarded in \cite{glp}. Below we construct
a simple unitary transformation which  relates a quantum mechanics with
the
conformal Galilean symmetry to a set of decoupled particles for which the same symmetry
is realized in a nonlocal way. Possible applications of the map are discussed.

Consider a representation of  the conformal Galilei algebra realized in a
non--relativistic quantum mechanics of $N$ particles in $d$--dimensional
space\footnote{We work in the
standard coordinate representation
$[x_\a^i, p_\b^j]=i \d^{ij} \d_{\a\b}$ with
$p_\a^i=-i \frac{\partial}{\partial x_\a^i}$ and use the natural units for which $\hbar=1$.}
\bea\label{gen}
&&
H=\sum_{\a=1}^N \sum_{i=1}^d \frac{1}{2 m_\a} p_\a^i p_\a^i+V(x_1, \dots, x_N):=H_0+V(x)\ ,
\quad  \qquad ~ \quad ~  P^i=\sum_{\a=1}^N p_\a^i\ ,
\nonumber\\[2pt]
&&
M^{ij}= \sum_{\a=1}^N(x_\a^i p_\a^j- x_\a^j p_\a^i)\ ,\quad  \quad \quad
K^i=\sum_{\a=1}^N m_\a x_\a^i\ -t P^i\ , \quad \quad \quad ~
 M=\sum_{\a=1}^N m_\a\ ,
\nonumber
\eea
\bea
&&
C=-t^2 H+2t D+\frac{1}{2} \sum_{\a=1}^N \sum_{i=1}^d m_\a x_\a^i x_\a^i\ , \qquad
D=tH-\frac{1}{4}\sum_{\a=1}^N \sum_{i=1}^d (x_\a^i p_\a^i+p_\a^i x_\a^i)\ .
\eea
Here $H$ is the Hamiltonian which generates time translation,
$P^i$, $M^{ij}$ and $K^i$ are the
generators of space translations, rotations and Galilei boosts, respectively.
$M$ is the mass operator (the central charge) and $C$ and $D$ are the generators of
special conformal transformation and dilatation, respectively. In the equations above
$i$ is the spatial index and $\a$ labels the distinct particles, with
$m_\a$ being the particle mass.

The commutation relations of the conformal Galilei algebra (vanishing commutators are
omitted)
\bea
&&
[M^{ij},M^{kl}]=i(\d^{ik} M^{jl}+\d^{jl} M^{ik}-\d^{jk} M^{il}-\d^{il} M^{jk})\ , \qquad
[M^{ij},P^k]=i(\d^{ik} P^j-\d^{jk} P^i)\ ,
\nonumber\\[2pt]
&&
[M^{ij},K^k]=i(\d^{ik} K^j-\d^{jk} K^i)\ ,~~ [K^i,P^j]=i\d^{ij} M\ , ~ ~
[H,K^i]=-i P^i\ , ~~ [D,K^i]=\frac{i}{2} K^i\ ,
\nonumber\\[2pt]
&&
[C,P^i]=i K^i\ ,\quad [D,P^i]=-\frac{i}{2} P^i\ , \quad
[H,D]=i H\ , \quad [H,C]=2i D\ , \quad [D,C]=i C
\eea
constrain the potential $V(x)$ which enters the Hamiltonian
to obey the system of partial differential equations
\bea\label{str}
&&
\sum_{\a=1}^N (x_\a^i \frac{\partial}{\partial x_\a^j}-x_\a^j
\frac{\partial}{\partial x_\a^i})V(x)=0\ ,\quad
\sum_{\a=1}^N \frac{\partial V(x)}{\partial x_\a^i}=0\ ,\quad
\sum_{\a=1}^N \sum_{i=1}^d x_\a^i \frac{\partial V(x)}{\partial x_\a^i}+2 V(x)=0\ .
\nonumber\\[2pt]
&&
\eea
Any solution to (\ref{str}) defines a quantum mechanics with the Galilean conformal symmetry.

The simplest solution $V(x)=0$ corresponds to a system of decoupled
particles which is governed by the free Hamiltonain $H_0$. Notice that
setting $t=0$ in (\ref{gen})
one does not spoil the algebra. It is this case which was considered
previously for one--dimensional systems in \cite{glp}. Below we take into
account the time dependent parts of the generators as well.
In what follows we shall use the notation
\be
C_0=\frac{1}{2} \sum_{\a=1}^N \sum_{i=1}^d m_\a x_\a^i x_\a^i\ , \qquad
D_0=-\frac{1}{4}\sum_{\a=1}^N \sum_{i=1}^d (x_\a^i p_\a^i+p_\a^i x_\a^i)\ .
\ee
The operators ($H$, $D_0$, $C_0$) obey the commutation relations of $so(1,2)$
provided the rightmost equation in (\ref{str}) holds.

As the next step consider an automorphism of the Galilei
algebra
\be\label{ser}
T \quad \rightarrow \quad T'=e^{iA}~ T~ e^{-iA}=T+\sum_{n=1}^\infty\frac{i^n}{n!}
\underbrace{[A,[A, \dots [A,T] \dots]}_{n~\rm times} \ ,
\ee
generated by a specific combination of the operators $H$, $D_0$, $C_0$
\be\label{a}
A\=a H + \frac{1}{a} C_0 -2D_0\ .
\ee
Here $a$ is an arbitrary parameter of the dimension of length and the number
coefficients are adjusted so as to terminate the infinite
series in (\ref{ser}) at a final step. Double commutators
of the generators of the Galilei algebra with the operator
$A$ prove to be proportional to
$A$ or vanish identically which automatically terminates the series.

Under the unitary transformation (\ref{ser}) the operators $M^{ij}$ and $M$ are unchanged.
For $P^i$ and $K^i$ one finds
\be
{P'}{}^i=-\frac{1}{a} (K^i+t P^i)\, \qquad K'{}^i=\left(2+\frac{t}{a}\right)K^i+
a{\left(1+\frac{t}{a}\right)}^2 P^i\ ,
\ee
while for the conformal generators a straightforward computation gives
\bea
&&
H'=\frac{1}{a^2} C_0\ ,\qquad D'=-D_0+\frac{1}{a}\left(2+\frac{t}{a} \right) C_0\ ,\nonumber\\[2pt]
&&
C'=a^2 (H_0+V(x))-2a\left(2+\frac{t}{a} \right)D_0+{\left(2+\frac{t}{a} \right)}^2 C_0 \ .
\eea
A remarkable fact is that
the Hamiltonian of the original interacting system is mapped to $C_0$, while the interaction
potential $V(x)$ is moved to $C'$.

Our next objective is to realize the map $C_0  \rightarrow H_0$ by applying the second
automorphism similar to (\ref{ser}), (\ref{a}). It turns out that the map
found in \cite{glp} for one--dimensional systems fits the higher dimensional case as well.
It is straightforward to verify that under the unitary transformation
\be\label{sec}
T' \quad \rightarrow \quad T^{''}=e^{iB}~ T'~ e^{-iB}\
\ee
with
\be
B=-a H_0-\frac{1}{a} C_0+2 D_0\
\ee
the operators $M^{ij}$ and $M$ remain inert, while ${P'}{}^i$,
${K'}{}^i$ regain their original form
\be
{P''}{}^i=P^i\ , \qquad {K''}{}^i=K^i\ .
\ee
For the conformal subalgebra the power series terminates at the third step and
after some calculation one gets
\be
H''=H_0\, \qquad D''=tH_0+D_0\ , \qquad
C''=-t^2 H_0+2t D''+C_0+a^2 \left( e^{iB}~V(x)~ e^{-iB}\right)\ .
\ee
But for the last term in $C''$, this is a representation of the conformal Galilei
algebra on a set of $N$ decoupled particles in $d$ dimensions.

Thus, the unitary operator $e^{i B} e^{i A}$ enables one to remove the potential
term $V(x)$ from the Hamiltonian which, however, resurfaces in the generator
of special conformal transformation as a {\it nonlocal} contribution. Hence,
the price paid for the simplification in the dynamics
is a nonlocal realization of the full conformal Galilei algebra in the
Hilbert space of a quantum mechanical system. Notice, that consistency
requires  $e^{i B} e^{i A}$ to be independent of the dimensionful
parameter $a$. That $\frac{d}{d a} (e^{i B} e^{i A})=0$ can be verified by
a straightforward calculation (see \cite{glp} for more details).

Finally, let us discuss possible applications of the transformation considered in this work.
Firstly, with appropriate modifications the transformation can be used for constructing
exactly solvable quantum mechanics models and
building the complete set of eigenstates as, for example, in \cite{gur,gpp}. Secondly, as
a set of decoupled particle is straightforward to supersymmetrize, the map
provides an efficient
means for constructing various supersymmetric many--body models in the spirit
of \cite{glp,glp1,glp2,ol,ser}. Thirdly, with some modification the mapping can be applied
in the context of trapped Fermi gases at unitarity (see \cite{mehen} and references therein).
Forthly, since the construction is purely algebraic,
it is likely to be applicable within the framework of
a nonrelativistic field theory as well.

\vspace{0.5cm}

\noindent{\bf Acknowledgments}\\

\noindent
We thank P. Horv\'athy and T. Mehen for useful comments.
The author is grateful to the Institut f\"ur Theoretische Physik at the Leibniz Universit\"at Hannover for the hospitality extended to him at the final stage
of this work. The research was supported by RF Presidential grants
MD-2590.2008.2, NS-2553.2008.2 and RFBR grant 06-02-16346.

\end{document}